\documentclass[11pt,twoside]{article}

\usepackage{asp2006}
\usepackage{epsf}
\usepackage{psfig}
\usepackage{lscape}

\begin{document}
\title{Distance to U Pegasi by the DDE Algorithm}   
\author{T. R. Vaccaro}   
\affil{Francis Marion University, Dept. of Physics \& Astronomy,P.O. Box 100547, Florence, SC 29502,USA}    
\author{Dirk Terrell}   
\affil{Southwest Research Institute, Dept. of Space Studies, Boulder, CO 80301, USA}    
\author{R. E. Wilson}   
\affil{University of Florida, Dept. of Astronomy, Bryant Space Science Center, P.O. Box 112055, Gainesville, FL 32611-2055,USA}    

\begin{abstract} 
A distance is found for the W UMa type binary U Pegasi, with a newly modified version of the Wilson-Devinney program (W-D) that makes use of the direct distance estimation (DDE) algorithm. The reported distance of $d=123.6$pc is an average based on solutions for $B$ and $V$ data and a primary star temperature of $5800$K. Standardized light curves (not differential), radial velocities, and a spectroscopic primary star temperature are input to the program.  Differential corrections were performed for each light curve band along with the velocities for two primary temperatures that span $100$K. $Log_{10}d$ is a model parameter like many others that are adjustable in W-D.  The eclipsing binary distance agrees with the \textit{Hipparcos} parallax distance and is more precise.
\end{abstract}


\section{Analysis}   
Light curves of the W-type W UMa eclipsing binary U Pegasi by \citet{zhai1984} were converted from differential magnitudes to the standard $B$ and $V$ systems by addition of the comparison star's magnitudes, $B=10^{m}.092 \pm 0^{m}.026$, $V=9^{m}.488 \pm 0^{m}.021$ \citep{kharchenko2001} for BD+14$^\circ$5078. Both U Peg components are of type G2 \citep{lu1985}, and an adopted primary star temperature of 5800K was rounded from two values in \citet{cox00}. Subsequent work by \citet{zhai1988} refined several velocity points that occur near conjunction. The light/velocity data were solved simultaneously for distance and other 
parameters by a version of the Wilson-Devinney (W-D) Differential Corrections (DC) program that contains the direct distance estimation (DDE) algorithm.

The \citet{zhai1988} parameters were initial input. However the linear ephemeris \citep{zhai1984} was independently adjusted to improve fits to $V$ band light curves and velocity curves, which are separated by 6 years.  A solution with primary temperature 5900K was also made to allow parameter interpolation in $T_1$ by readers. Adjusted parameters were  inclination ($i$), secondary temperature ($T_2$), semi-major axis ($a$), potential ($\Omega_1$), systemic velocity ($V_{\gamma}$), mass ratio ($q=\frac{M_2}{M_1}$), and distance ($log_{10} d$).  Initial solutions that adjusted third light found $\ell_3$ to be insignificant, so $\ell_3$ was fixed at zero. The parameters were split into two DC subsets to reach the solutions of Table 1.

\section{Final Remarks}   
The theory of the DDE algorithm, with overviews of simulations, is in \citet{wilson2008}.
As applied to 1978 Zhai, et al. light curves and 1984 Lu velocity data for U Peg, DDE yields respective $B,V$ distances of 121.6 and 125.5 pc if $T_1=5800$K. The formal errors are smaller than the spread, so calibration errors probably account for most of the modest disagreement.  Hipparcos parallaxes, as revised by \citet{van Leeuwen2007}, span 122 to 175 pc according to the reported $1\sigma$ errors. Our model light curves might fit the data better if spots were modeled as in \citet{djurasevic2001}, but derived distance typically is not much affected by such refinement. Although period variations have been linked to a third body by several authors, our solutions indicate that $\ell_3$ is unlikely to affect distance estimates significantly.  Another set of distances will be based on light curves by T. Pribulla \citep{pribulla2002} and published elsewhere.

This work was supported in part by NSF grant AST-0307561 at the University of Florida.
\begin{table}[!ht]
\caption{Solutions for U Peg}
\begin{center}
{\small
\begin{tabular}{ccccc}
\tableline
Parameter & $B$($T_1$= 5800K) & $V$($T_1$= 5800K) & $B$($T_1$= 5900K) & $V$($T_1$= 5900K) \\
\tableline
$i$[$\deg$] & $75.26 \pm 0.08$ & $75.44 \pm 0.08$ & $74.92 \pm 0.09$ & $75.35 \pm 0.12$ \\
$T_2$[K] & $5637 \pm 4$ & $5622 \pm 8$ & $5738 \pm 8$ & $ 5619 \pm 7$ \\
$a[R_{\sun}]$ & $2.56 \pm 0.02$ & $2.54 \pm 0.02$ & $2.57 \pm  0.02$ & $ 2.56 \pm  0.02$ \\
$V_{\gamma}$ [km s$^{-1}]$ & $30.9 \pm 0.8$ & $30.7 \pm 1.0$ & $31.0 \pm 0.9$ & $ 31.2 \pm 0.9$ \\
$\Omega_{1}$ & $6.294 \pm 0.012$ & $6.207 \pm 0.025$ & $6.256 \pm 0.009$ & $ 6.201 \pm 0.013$ \\
$q = M_{2}/M_{1}$ & $2.839 \pm 0.009$ & $2.763 \pm 0.019$ & $2.824 \pm 0.006$ & $ 2.772 \pm 0.010$ \\
$d$[pc] & $121.6 \pm 1.0$ & $125.5 \pm 1.0$ & $129.3 \pm 1.0$ & $ 132.2 \pm 1.0$ \\
$R_{1}/a$ (pole) & $0.2809 \pm 0.0011$ & $0.2819 \pm 0.0023$ & $0.2826 \pm 0.0008$ & $ 0.2830 \pm 0.0012$ \\
$R_{1}/a$ (side) & $0.2938 \pm 0.0013$ & $0.2947 \pm 0.0028$ & $0.2958 \pm 0.0010$ & $ 0.2961 \pm 0.0015$ \\
$R_{1}/a$ (back) & $0.3327 \pm 0.0050$ & $0.3327 \pm 0.0050$ & $0.3358 \pm 0.0018$ & $ 0.3352 \pm 0.0027$ \\
$R_{2}/a$ (pole) & $0.4506 \pm 0.0008$ & $0.4474 \pm 0.0018$ & $0.4514 \pm 0.0006$ & $ 0.4490 \pm 0.0010$ \\
$R_{2}/a$ (side) & $0.4483 \pm 0.0012$ & $0.4803 \pm 0.0025$ & $0.4854 \pm 0.0009$ & $ 0.4824 \pm 0.0013$ \\
$R_{2}/a$ (back) & $0.5131 \pm 0.0015$ & $0.5089 \pm 0.0032$ & $0.5148 \pm 0.0011$ & $ 0.5114 \pm 0.0017$ \\
\tableline
\end{tabular}
}
\end{center}
\end{table}

\end{document}